\documentclass[twocolumn,showpacs,preprintnumbers,amsmath,amssymb]{revtex4}
\usepackage{dcolumn}% Align table columns on decimal point
\usepackage{bm}% bold math
\usepackage{latexsym}
\usepackage{amsfonts}
\usepackage[dvips]{graphicx}
\usepackage[usenames]{color}
\usepackage{makeidx}
\newcommand{\e}{\epsilon}

\newcommand{\ket}{\rangle }
\newcommand{\bra}{\langle }

\newcommand{\up}{\uparrow}
\newcommand{\dw}{\downarrow}

\newcommand{\tl}[1]{\tilde{#1}}
\newcommand{\tP}{{\tl{P}}}
\newcommand{\tQ}{{\tl{Q}}}
\newcommand{\tR}{{\tl{R}}}
\newcommand{\tS}{{\tl{S}}}

\newcommand{\tB}{{\tl{B}}}
\newcommand{\tA}{{\tl{A}}}

\newcommand{\tq}{{\tl{q}}}
\newcommand{\trr}{{\tl{r}}}

\newcommand{\tc}{{\tl{c}}}

\begin{document}

%\preprint{APS/123-QED}

\title{Breakdown of an Electric-Field Driven System:\\
a Mapping to a Quantum Walk}
\author{Takashi Oka, Norio Konno$^\dagger$, Ryotaro Arita$^*$ and Hideo Aoki}
\address{Department of Physics, University of Tokyo, Tokyo 113-0033, 
Japan\\
$\dagger$Department of Applied Mathematics, Faculty of Engineering, Yokohama National University,
 Yokohama 240-8501, Japan}

\date{\today}
\begin{abstract}
\noindent 
Quantum transport properties of electron systems 
driven by strong electric fields are 
studied by mapping the Landau-Zener 
transition dynamics to a 
quantum walk on a semi-infinite one-dimensional lattice
with a reflecting boundary,
where the sites correspond to energy levels
and the boundary the ground state.
Quantum interference induces a distribution 
localized around the ground state, 
and a delocalization transition occurs when 
the electric field is increased, which describes 
the dielectric breakdown in the original electron system.
\end{abstract}
\maketitle

%%%%%%%%%%%%%%%%%%%%%%%%%%%%%%%%%%%%%%%%%%%%%%%%%%
%{\it Introduction}
%%%%%%%%%%%%%%%%%%%%%%%%%%%%%%%%%%%%%%%%%%%%%%%%%%
Dynamics of quantum statistical systems driven by finite external fields 
has attracted much attention as a typical class of non-equilibrium 
phenomenon.  One problem is the ``dissipation'' arising in  
electron systems driven out of their ground states by 
strong electric fields as studied by many authors
\cite{Oka2003,Lenstra1986,
Landauer1985,Gefan1987,Blatter1988,Cohen2000PRL,Fishman1982}. 
In these references the electric field is expressed 
(via Faraday's electro-magnetic induction) as a time-dependent Aharonov-Bohm(AB) 
flux, and this induces, for strong fields, interlevel 
Landau-Zener transitions.  
There, an issue is whether the bunch of energy levels 
around the one having the main amplitude 
can play a role akin to dissipation\cite{Cohen2000PRL}.  
The purpose of the present study is to propose a mapping of the 
system onto a quantum algorism model for studying the problem.  
We start by noting that the driven quantum system and 
the quantum algorism model have in fact similarities: 
The former deals with quantum transitions among 
macroscopic number of energy levels 
while the latter describes the dynamics of many qubits governed
by quantum logic gates, for which powerful analytical techniques are 
being developed. 

In the one-dimensional tight-binding model, 
the Hamiltonian is 
$
H(\phi(t))=-t\sum_{i,\sigma}\left[e^{-i\frac{2\pi}{L}\phi(t)}c_{i+1\sigma}^\dagger 
c_{i\sigma}+\mbox{h.c.}\right]+\hat{V}$,
where $c_{i,\sigma}^\dagger$ creates an electron 
at $i=1,2,\ldots L$ with spin $\sigma=\up,\dw$, 
the AB-flux measured in units of the flux quantum 
$\phi(t)=FLt/h$ represents the electric field $F$, and 
$\hat{V}$ the external potential or electron-electron interaction. 
Disordered mesoscopic 
systems\cite{Lenstra1986,Landauer1985,Gefan1987,Blatter1988}
have been treated with this Hamiltonian, which has been 
extended to a strongly correlated electron 
system by three of the present authors\cite{Oka2003}.  
The adiabatic energy levels against $\phi$ have 
many level anti-crossings (as schematically shown in Fig.1(a)), 
which come from the disorder in disordered (one-body) systems, or 
from Umklapp processes in correlated systems. 
The system obeys 
the time-dependent Schr\"odinger's equation 
starting from the ground state $|\Psi_0\ket$ of $H(0)$,
where its formal solution is
$\hat{T}e^{-i\int_0^tH(\phi(s))ds}|\Psi_0\ket$. 
When the field is finite, non-adiabatic Landau-Zener tunneling processes 
occur at the level anti-crossings, which take place 
first from the ground state and then among higher excited states.
Authors of \cite{Lenstra1986,Gefan1987,Blatter1988} 
introduced a transfer matrix representation to 
mimic the evolution $\hat{T}e^{-i\int_0^tH(\phi(s))ds}$, where 
a set of $2\times 2$ unitary matrices 
represent the transitions among neighboring energy levels.  
Extensive studies on this model have found the existence of 
a {\it dynamical localization}\cite{Fishman1982}, or 
an Anderson localization in energy space\cite{Blatter1988}. 
However, the effect of quantum interference, which should be 
the key to the Anderson localization, has been studied only 
numerically\cite{Landauer1985,Lenstra1986,Gefan1987,Blatter1988}.

%%%%%%%%%%%%%%%%%%%%%%%%%%%%%%%%%%%%%%%%%%%
\begin{figure}[t]
\centering 
\includegraphics[width=8.cm]{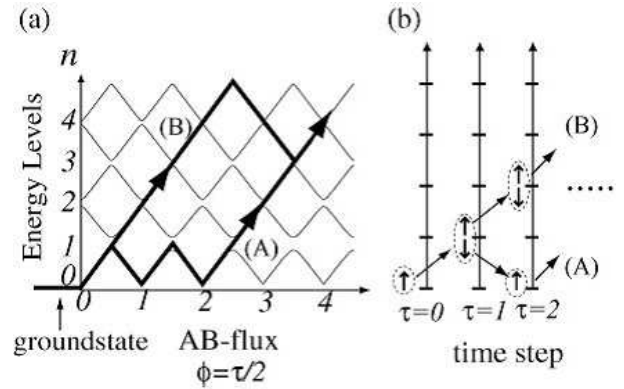}
\caption{
(a) Idealized energy levels of 
an electron system on a ring 
plotted against the AB-flux 
$\phi=FLt/h$, which increases linearly with time 
after the electric field $F$ is turned on at $t=0$.
Two paths (A, B) for the Landau-Zener transition among neighboring 
levels are shown.  
(b) A mapping to a quantum-walk, where the 
energy levels are mapped to sites of the qubits and the 
ground state to a reflecting boundary.
}
\label{LZ2QW}
\end{figure}
%%%%%%%%%%%%%%%%%%%%%%%%%%%%%%%%%%%%%%%%%%%

Now we want to point out that 
the Landau-Zener transitions in multi-level systems 
presented above 
can be mapped to a {\it quantum walk} on a lattice.  
We shall show that the localization (in energy axis) of wave functions 
can be studied in terms of an exact solution for the 
transition amplitudes in the quantum walk.
Quantum walk is a quantum counterpart of the 
random walk\cite{Kempe2003,Tregenna2003,ChalkerCoddington}).  
The mapping we conceive here between the quantum walk 
and the Landau-Zener dynamics is straightforward: 
A qubit on site $n\ge 0$ labeled by 
$L,R$ corresponds to the states at the $n$-th anticrossing point,
which moves down ($L$; with a left-going current) 
or up ($R$; right-going) in energy after the 
tunneling event(Fig.\ref{LZ2QW}).  
One important point we note here is that the 
mapped quantum walk has a {\it reflecting} boundary, 
since we cannot walk below the ground state in energy.  
In previous quantum-walk 
studies\cite{Bach2002absorption,Yamasaki2003,Konno2003} 
an absorbing boundary was considered, for which the
generating functions were obtained\cite{Konno2003}.  
Here we first obtain the generating function for the quantum walk 
with a reflecting boundary.
We have found the 
existence, and an analytic form for, 
the amplitude localized around the boundary (the ground state 
in the original problem). 
The solution exhibits an asymmetry between the 
L and R states, which represents  
a finite total momentum.  When the electric field exceeds a 
critical value, a delocalization transition (on energy axis) 
is observed, which we identify here to describe the 
dielectric breakdown in the original electron model.

%%%%%%%%%%%%%%%%%%%%%%%%%%%%%%%%%%%%%%%%%%%%%%%%%%
{\it Mapping ---}
%%%%%%%%%%%%%%%%%%%%%%%%%%%%%%%%%%%%%%%%%%%%%%%%%%
Let the wave function for the $n$-th energy level 
at time $\tau$ (measured in units of 
$\Delta t/2=h/2FL$) be $\Psi(n,\tau)= \vspace{0cm}^t(
	\psi_L(n,\tau),
	\psi_R(n,\tau))$,
where $\psi_{L}(n,\tau) (\psi_{R}(n,\tau))$ 
has a left- (right-) momentum. 
Each energy level is subject to a Landau-Zener tunneling 
(with certain probability and phase) to neighboring levels in 
a time period $\Delta t/2$. 
While transitions among more than 
three levels exist in principle, 
here we restrict ourselves to transitions
among neighboring levels, for which 
the quantum tunneling can be described by a set of 
$2\times 2$ unitary 
matrices\cite{Lenstra1986,Landauer1985,Gefan1987,Blatter1988}.
The diagonal elements represent Landau-Zener 
tunneling, while the off-diagonal ones 
the backward scattering process. 
Here we simplify the  problem by assuming that the matrices are 
the same $U=\left(
	\begin{array}{cc}
	a&b\\
	c&d
	\end{array}
\right)$
except for the one 
at the boundary,
$\tilde{U}\equiv \left(
	\begin{array}{cc}
	\tilde{a}&\tilde{b}\\
	\tilde{c}&\tilde{d}
	\end{array}
\right).$
As we shall see, an interesting structure arises 
in the overall 
shape of the distribution in the bounded quantum walk 
even within this simplification.\cite{different}  
The wave function 
evolves deterministically 
following a recursion formula,
\begin{equation}
\Psi(n,\tau+1)=P\Psi(n+1,\tau)+Q\Psi(n-1,\tau)\label{BQW1}
\end{equation}
for excited ($n\ge 2$) states, while we put 
\begin{eqnarray}
\Psi(1,\tau+1)&=&P\Psi(2,\tau)+\tilde{Q}\Psi(0,\tau),\label{BQW2} \nonumber \\
\Psi(0,\tau+1)&=&P\Psi(1,\tau)\label{BQW3}
\end{eqnarray}
between the ground and the first excited levels.  Here we have 
decomposed $U$ into 
$P \equiv (^{ab}_{00})$, $Q \equiv (^{00}_{cd})$
and $\tilde{Q} \equiv (^{00}_{\tilde{c}\tilde{d}})$. 
Equations(\ref{BQW1}),(\ref{BQW3}) 
define the mapping to a one-dimensional quantum walk 
on a semi-infinite space with a reflecting boundary at $n=0$.
We take the initial wave function $|\Psi_0\ket$ 
to be the ground state, i.e., $\Psi(0,0)= \vspace{0cm}^t(1,0)$
and $\Psi(n,0)=0$ for $n\ge 1$.

%%%%%%%%%%%%%%%%%%%%%%%%%%%%%%%%%%%%%%%%%%%%%%%%%%
{\it Generating function for the quantum walk ---}
%%%%%%%%%%%%%%%%%%%%%%%%%%%%%%%%%%%%%%%%%%%%%%%%%%
Now we state our main result for the quantum walk: 
We can obtain the generating function for the wave function 
in the reflecting boundary condition, where 
the {\it generating function} is defined as 
$\Psi^{L,R}(0\to n;z)=\sum_{\tau =0}^\infty\psi_{L,R}(n,\tau)\;z^{\tau}$, 
with the summation taken over all the paths starting from $0$ and 
ending at $n$.  The obtained expression is 
\begin{eqnarray}
\Psi^{L}(0\to n;z)&=&\left(\frac{d\lambda_+(z)}{a}\right)^{n-1}
\frac{\tilde{c}d}{ac}\;\frac{\lambda_+(z)-az}{1-\tilde{c}A^r(0\to 0;z)}
\label{GF1},\nonumber \\
\Psi^{R}(0\to n;z)&=&\left(\frac{d\lambda_+(z)}{a}\right)^{n-1}
\frac{\tilde{c}z}{1-\tilde{c}A^r(0\to 0;z)},\nonumber \\
&&\hspace{-2.8cm}
\lambda_+(z) = \frac{\Delta z^2+1-\sqrt{\Delta^2z^4+2\Delta(1-2|a|^2)z^2+1}}{2dz}.
\end{eqnarray}
Here  $\Delta \equiv {\rm det}U =ad-bc$, and 
$A^r(0\to 0;z) = [d\lambda_+(z)-\Delta z]z/c$
is the generating function for the quantum walk with an 
absorbing boundary, governed by Eqs.(\ref{BQW1},\ref{BQW3}) 
with $\tQ=0$\cite{Konno2003}.  

%%%%%%%%%%%%%%%%%%%%%%%%%%%%%%%%%%%%%%%%%%%%%%%%%%
%{\it Derivation}
%%%%%%%%%%%%%%%%%%%%%%%%%%%%%%%%%%%%%%%%%%%%%%%%%%
In order to obtain the generating function (\ref{GF1}),
we first express the transition amplitudes 
$\tilde{\Xi}^b(0\to n;\tau)$ as a sum of the contributions
from  paths starting from the $0$-th level (ground state) 
ending at the $n$-th level after $\tau$ steps. 
For example, 
$\tilde{\Xi}^b(0\to 2;4)=QPQ\tilde{Q}+PQ^2\tilde{Q}+Q\tilde{Q}P\tilde{Q}$
consists of three paths, 
where the quantum nature of the walk appears as 
the matrices being complex, 
 i.e., different paths interfere 
with each other. 
The transition amplitude relates the 
initial vector $\Psi(0,0)$ 
to the wave function via $\Psi(n,\tau)=\tilde{\Xi}^b(0\to n;\tau)\Psi(0,0)$. 
We also introduce $\tilde{\Xi}^a(0\to n;\tau)$ as the 
transition amplitude in the quantum walk with an absorbing boundary;  
e.g., $\tilde{\Xi}^a(0\to 2;4)=QPQ\tilde{Q}+PQ^2\tilde{Q}$. 

We obtain the transition amplitudes by employing 
a technique called the PQRS method\cite{Konno2002a}.
Since the four 
matrices $P$, $Q$, $R\equiv(^{cd}_{00})$ and $S\equiv(^{00}_{ab})$,
with $\tP,\;\tQ,\;\tR,\;\tS$ defined similarly, 
span an orthonormal basis in $M_2(\Bbb{C})$
with an inner product $\bra A|B\ket={\rm Tr}(A^\dagger B)$,
we can express the transition amplitude $\tilde{\Xi}^{b}(0\to n;\tau)$
as a linear combination, 
\begin{equation}
\tilde{\Xi}^{b}(0\to n;\tau)=
\tilde{b}^\tq(0\to n;\tau)\tilde{Q}+\tilde{b}^\trr(0\to n;\tau)\tilde{R}\label{ExpansionPQRS}
\end{equation}
with the tilde basis, where $\tP$ and $\tS$ terms are absent.
Thus, eq.(\ref{ExpansionPQRS}) defines a set of 
complex numbers $\tilde{b}^\tq(0\to n;\tau)$ and $\tilde{b}^\trr(0\to n;\tau)$,
and we denote $\tB^{\tq,\trr}(0\to n;z)=
\sum_{\tau=0}^\infty\tilde{b}^{\tq,\trr}(0\to n;\tau)z^{\tau}$
as their generating functions.
The generating functions  
$\tA^{\tq,\trr}(0\to n;z)$ for the absorbing boundary problem is defined similarly
from $\tilde{\Xi}^{a}(0\to n;\tau)=
\tilde{a}^\tq(0\to n;\tau)\tilde{Q}+\tilde{a}^\trr(0\to n;\tau)\tilde{R}$.  
We note that for $n=0$ the $\tilde{Q}$ terms vanish, 
i.e., $\tB^{\tq}(0\to 0;z)=\tA^{\tq}(0\to 0;z)=0$.
For a special case $\tilde{U}=U$, we drop all the tilde 
marks $\Xi^{b}(0\to n;\tau),\;B^{q,r}(0\to n;z),\;b^{q,r}(0\to n;\tau)\ldots$.

The PQRS method establishes a connection between 
the generating functions via multiplication 
rules\cite{Konno2002a} for the $PQRS$ matrices.
From a recursion relation between $\tilde{\Xi}^b(0\to 0;\tau)$
and $\Xi^b(0\to n-1;\tau-1)$,
we obtain 
\begin{eqnarray}
\tB^\tq(0\to n;z)&=&\left[ 1+\tc\tB^\trr(0\to 0;z)\right]d zB^q(0\to n-1;z),
\label{RecursionRelation}\\
\tB^\trr(0\to n;z)&=&\left[ 1+\tc\tB^\trr(0\to 0;z)\right]d zB^r(0\to n-1;z).\nonumber
\end{eqnarray} 
With a repeated application of the no-tilde version of 
Eq.(\ref{RecursionRelation}) we obtain 
$B^q(0\to n;z)=[d\lambda_+(z)/a]^n/d,\; 
B^r(0\to n;z)=[d\lambda_+(z)/a]^n[(\lambda_+(z)-az)/(acz)]$.  
Again using (\ref{RecursionRelation})
with a relation between $\tB^{\trr}(0\to 0;z)$
and $A^r(0\to 0;z)$,
we arrive at eqs.(\ref{GF1}) after a bit of algebra.

%%%%%%%%%%%%%%%%%%%%%%%%%%%%%%%%%%%%%%%%%%%%%%%%%%
{\it Edge states in energy space ---}
%%%%%%%%%%%%%%%%%%%%%%%%%%%%%%%%%%%%%%%%%%%%%%%%%%
One asset of the generating function is that we can show that 
there exist ``edge states" in the bounded quantum walk.  
To actually show this let us take the Landau-Zener tunneling probability,
$
p=\exp(-\pi\overline{F}/F),
$
where the Zener threshold electric field 
$\overline{F}\propto \Delta^2$ 
depends on the size of the gap $\Delta$\cite{Zener1934,Landau}.
We plug this in the transfer matrix,
\begin{equation}
U=\left(
	\begin{array}{cc}
	\sqrt{p}e^{i\beta}&\sqrt{1-p}e^{i\gamma}\\
	-\sqrt{1-p}e^{-i\gamma}&\sqrt{p}e^{-i\beta}
	\end{array}
\right),\;\tilde{U}=\left(
	\begin{array}{cc}
	0&e^{i\tilde{\gamma}}\\
	-e^{-i\tilde{\gamma}}&0
	\end{array}
\right),
\end{equation}
where the diagonal elements of $\tilde{U}$ are 
set to zero to represent a complete reflection.
The phases $\beta,\gamma,\tilde{\gamma}$ consist of the dynamical phase 
and the geometric phase (nonadiabatic counterpart to Berry's phase).   
In physical systems (such as the Hubbard model) 
these phases at the level anti-crossings are generally nonzero 
(and depend on the driving force $F$).  

\begin{figure}[t]
\centering 
\includegraphics[width=8.5cm]{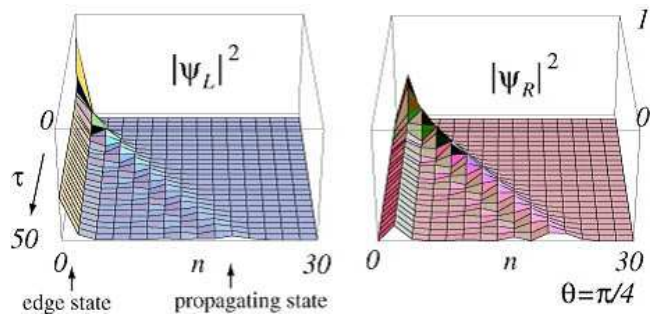}
\caption{
Time evolution of $|\psi_{L,R}(n,\tau)|^2$ with 
the initial condition $\psi_L(0,0)=1$. 
Here we set $p=0.2<p_c={\rm sin}^2\theta$, and 
only the amplitudes at even $n$ are displayed.
}
\label{prob}
\end{figure}

Expanding the generating function (\ref{GF1}) 
into a power series in $z$ yields the time evolution of the
wave function (Fig.\ref{prob}). 
We can see that $\Psi$ (distribution on energy axis) 
breaks up, after a transient period, into two parts, 
$
\Psi(n,\tau)\sim\Psi_{\rm edge}(n,\tau)+\Psi_{\rm bulk}(n,\tau)
$,
where $\Psi_{\rm edge}$ is a component localized around the 
boundary (i.e., the ground state) 
while $\Psi_{\rm bulk}$ is a component traveling into 
the bulk (excited states) with a nearly constant velocity. 
Interestingly, the edge state 
only appears when the phase difference,
\[
\theta\equiv\gamma-\tilde{\gamma},
\]
between the bulk and boundary transfer matrices is nonzero 
(as is the case with physical systems as mentioned).   

%%%%%%%%%%%%%%%%%%%%%%%%%%%%%%%%%%%%%%%%%%%
%%%%%%%%%%%%%%%%%%%%%%%%%%%%%%%%%%%%%%%%%%%
What is the  nature of the edge state?  The two generating functions 
$\Psi^{L,R}_{(0\to n)}(z)$ (eqs.(\ref{GF1}))
have a common, first-order pole in $z^2$ at 
$z_{\rm pole}^2=\frac{1-e^{-i\theta}\sqrt{1-p}}{1-e^{i\theta}\sqrt{1-p}}$. 
We can then obtain, with the Darboux's method\cite{Darboux}, 
the asymptotic wave function, 
%\begin{equation}
$
\Psi_{\rm edge}(n,\tau)\to \exp\left({-i\e \tau \Delta t/2\hbar}\right)\Phi_{\rm edge}(n), 
$
%\end{equation}
which obeys Floquet's theorem\cite{FloquetReview} with the Floquet 
mode $\Phi$ and the Floquet quasi-energy per length
%\begin{equation}
$
\e/L = (F/2\pi)\arg z_{\rm pole}^2,
$
%\end{equation}
a function of the electric field $F$ 
since $p=\exp(-\pi\overline{F}/F)$ involves $F$.
We note that $\e$ can be expressed as 
$\e=-i\hbar \lim_{t\to \infty}\frac{1}{t }\ln\bra \Psi_0|U(t)|\Psi_0\ket$,
i.e., the asymptotic effective Hamiltonian of the system. 
An important observation here is that 
the elements of the Floquet state $\Phi_{\rm edge}(n)= \vspace{0cm}^t(\phi^L(n),\phi^R(n))$
with
$\phi^{L,R}(n)=\lim_{z^2\to z^2_{\rm pole}}\Psi^{L,R}(0\to n;z)(1-z^2/z^2_{\rm pole})$, 
%given by
%\begin{eqnarray}
%\phi^L(k) = \left(\frac{\sqrt{p}e^{-i\phi}\bar{z}}{1-e^{-i\theta}\sqrt{1-p}}
%\right)^{k}\frac{2\sqrt{1-p}(\sqrt{1-p}-\cos\theta)}{\bar{z}^{2}(1-e^{i\theta}
%\sqrt{1-p})^2}
%\end{eqnarray}
%or $k\ge 0$ and
%\begin{eqnarray}
%\phi^R(k)&=&
%-\left(\frac{\sqrt{p}e^{-i\phi}\bar{z}}{1-e^{-i\theta}
%\sqrt{1-p}}\right)^{k-1}\\
%&&\hspace{1cm}\times\frac{2\sqrt{1-p}(\sqrt{1-p}-\cos\theta)
%e^{-i\tilde{\gamma}}}{\bar{z}(1-e^{i\theta}\sqrt{1-p})^2}\nonumber
%\end{eqnarray}
%for $k\ge 2$ with $\phi^R(0)=0$.
%The amplitudes 
form a {\it geometric series}, 
\begin{eqnarray}
|\phi_L(n)|^2 &=&r^{n}\;|\phi_L(0)|^2,\label{FloquetAmplitude}
 \nonumber \\
|\phi_R(n\ge 2)|^2 &=& r^{n-1}\;|\phi_L(0)|^2,\quad n=0,2,\ldots,  \nonumber \\
r &=& p/(2-p-2\cos\theta\sqrt{1-p}),
\end{eqnarray}
with $|\phi_L(0)|^2=(1-r)^2$.
Thus $\Phi(n)$ is an edge state exponentially 
localized (on energy axis) around the boundary $n=0$,
whose weight is $|\Phi_{\rm edge}|^2=\sum_n(|\phi_L(n)|^2+|\phi_R(n)|^2)=1-r$.
The size (on energy axis) of the edge state is
$
\xi=1/|\ln r|,
$
which behaves as $\xi\propto F/\overline{F}$ in the
small $F$ regime and diverges like 
$\xi\propto \;|p-p_c(\theta)|^{-1} \propto 
\overline{F}|F-F_c(\theta)|^{-1}$ 
in the vicinity of the threshold, 
$p_c(\theta)=\sin^2\theta, 
F_c(\theta)=-\pi\overline{F}/(2\ln\sin\theta$). 
When $p$ exceeds $p_c(\theta)$, the edge state 
collapses, and only the component propagating 
into the bulk remains.

%%%%%%%%%%%%%%%%%%%%%%%%%%%%%%%%%%%%%%%%%%%%%%%%%%
{\it Translation to electron systems ---}
%%%%%%%%%%%%%%%%%%%%%%%%%%%%%%%%%%%%%%%%%%%%%%%%%%
Having presented the results for the quantum walk, we are 
now in position to translate them back to the electron system. 
The electric field $F$ causes a production of electron-hole pairs 
through the Landau-Zener tunneling. 
However, these excited charges cannot be accelerated 
indefinitely, since the electron-electron interaction or disorder 
scatter the momenta due to the ``back scattering" at the 
level anti-crossings.  
This leads to bifurcations of the amplitude 
in energy space (whose example is depicted in Fig.\ref{LZ2QW}(a)).  
When different paths meet, quantum interference induces 
localization of the wave function, and 
the edge state $|\phi_{L,R}(n)|^2$, occurring for $\theta\neq 0$, 
in the quantum walk problem 
represents an effect of such an interference. 
So the present result is an analytic version 
of the edge states observed numerically 
in the long-time limit\cite{Fishman1982,Gefan1987,Blatter1988}. 

%%%%%%%%%%%%%%%%%%%%%%%%%%%%%%%%%%%%%%%%%%%
\begin{figure}[htbp]
\centering 
\includegraphics[width=7.5cm]{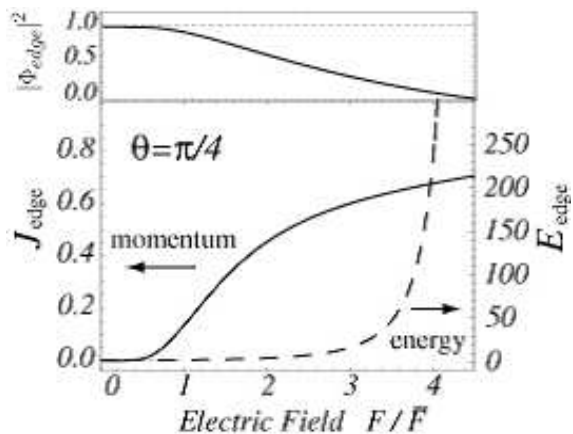}
\caption{
Total momentum $J_{\rm edge}$ and energy $E_{\rm edge}$ 
contributed by the edge component (eq.(\ref{J5E})) 
against the electric field $F$ for $\theta=\pi/4$. 
The upper figure depicts the weight of the edge state.
}
\label{crntene}
\end{figure}
%%%%%%%%%%%%%%%%%%%%%%%%%%%%%%%%%%%%%%%%%%%

The momentum and energy expectation values (summed over L and R 
states) for the edge state are
\begin{eqnarray}
J_{\rm edge}&=&\sum_{n}\left[j^R_n|\phi_R(n)|^2+j^L_n|\phi_L(n)|^2 \right]/|\Phi_{\rm edge}|^2,
\nonumber \\
E_{\rm edge}&=&\sum_{n}\left[E^R_n|\phi_R(n)|^2+E^L_n|\phi_L(n)|^2 \right]/|\Phi_{\rm edge}|^2.
\label{J5E}
\end{eqnarray}
Here we normalize them by 
the total amplitude of the edge state $|\Phi_{\rm edge}(n)|^2$, 
which decreases as we increase the electric field
and becomes zero at $F=F_c(\theta)$ (top panel of Fig.\ref{crntene}). 
Following \cite{Lenstra1986,Landauer1985,Gefan1987,Blatter1988},
we choose $j^{R,L}_n=\pm j_0n,\; E^{R,L}_n=E_0n^2,\;n=0,2,\ldots$
as the momentum and energy of the states, respectively,  
where $j_0,\; E_0$ are units of momentum and energy. 
We note that the momentum measures the chiral asymmetry (between 
$L$ and $R$) of the distribution.
By plugging in eq.(\ref{FloquetAmplitude}) we obtain 
$J_{\rm edge}=j_0\frac{p(2-p-2\cos\theta\sqrt{1-p})}{2\sqrt{1-p}(\sqrt{1-p}\cos\theta-1)^2}$, 
and a similar expression for $E_{\rm edge}$, valid 
for $F<F_c(\theta)$ (Fig.\ref{crntene}).  
In the weak-field regime, the $J_{\rm edge}$ is suppressed 
until the Landau-Zener tunneling 
is activated for $F > \overline{F}$ when 
$J_{\rm edge} \propto \exp(-\pi\overline{F}/F)$ starts to rise. 
When $F$ is further increased to reach $F=F_c(\theta)$, 
the electric-field induced breakdown
occurs, which corresponds, in the present picture, 
to the edge-to-propagating transition in the quantum walk. 
At the breakdown point the energy expectation value diverges as 
$E_{\rm edge}\sim |F-F_c(\theta )|^{-2}$. 
The $J_{\rm edge}$, however, does not diverge but shows a
smooth increase.

%%%%%%%%%%%%%%%%%%%%%%%%%%%%%%%%%%%%%%%%%%%%%%%%%%
{\it Discussions ---}
%%%%%%%%%%%%%%%%%%%%%%%%%%%%%%%%%%%%%%%%%%%%%%%%%%
The idea of mapping the non-equilibrium problem to 
a quantum walk may apply to wider 
range of systems having many energy gaps. 
Here we have concentrated on the ``edge state" when 
there is randomness (i.e., $\theta\neq 0$) only at the edge.   
Randomness in the transfer matrices in the excited state, ignored here, 
is expected to enhance the localization 
to bring $|\Phi_{edge}|^2$ closer to unity. 
Blatter {\it et.al}\cite{Blatter1988} 
have in fact obtained the expectation value of the momentum in such a situation, 
where the result resembles $J_{edge}$ obtained here. 
However, Blatter {\it et.al} did not obtain, 
in the range of electric field $F$ they have studied, 
a delocalization transition. 
In fact, Cohen {\it et.al}, in a random matrix model, 
found a disappearance of 
localized states in strong external fields\cite{Cohen2000PRL}.
If such transitions also occur in electron systems, 
the quantum-walk picture may be used in understanding the transition. 
Indeed, some of the quantum-walk models now under study 
exhibit localization (e.g., \cite{Inui2004multi}), 
and the present mapping may shed lights on the 
non-equilibrium properties of driven quantum systems.

%%%%%%%%%%%%%%%%%%%%%%%%%%%%%%%%%%%%%%%%%%%%%%%%%%
%{\it Acknowledge:}
%%%%%%%%%%%%%%%%%%%%%%%%%%%%%%%%%%%%%%%%%%%%%%%%%%
It is a pleasure to acknowledge V. Kagalovsky, Y. Matsuo, S. Miyashita, 
N. Nagaosa, and Y. Tokura for 
fruitful discussions in the early stage of the present work. 
We also wish to thank M. Katori and T. Sasamoto 
who suggested a link between the Landau-Zener transition and quantum walks. 
One of us (TO) thanks the Yukawa Institute for Theoretical Physics at 
Kyoto University, where this work was initiated
during the YITP-W-03-18 on ``Stochastic models in statistical mechanics".
Part of this work was supported by a JSPS fellowship for Young Scientists.

$^*$ Present address:
Max-Planck-Institut f\"{u}r Festk\"{o}rperforschung, Stuttgart, D-70569 Germany
%\bibliographystyle{prsty.bst}
%\bibliography{ref.bib}

\begin{thebibliography}{10}
\bibitem{different} 
When the evolution matrix $U$ (including the phase $\gamma$) 
differs from one level crossing 
to another, we can still use eq.(\ref{RecursionRelation}) to recursively obtain 
the generating functions, which will be discussed in a separate publication.
\bibitem{Darboux}
Darboux's method states that, if $F(t)=\sum_n F_n t^n$ 
can be expressed as $F(t)=\frac{f(t)}{(1-\gamma t)^\alpha}$
with a non-singular function $f(t)$, then the $n\to \infty$ then the $n\to \infty$ 
asymptote become 
$F_n = \frac{\gamma^n n^{\alpha-1}}{\Gamma(\alpha)}f(\gamma^{-1})(1+O(1/n))$.
\end{thebibliography}
%\printindex

\end{document}